# DECAY OSCILLATIONS IN ELECTRON CAPTURE AND THE NEUTRINO MASS DIFFERENCE

Murray Peshkin

*Physics Division, Argonne National Laboratory, Argonne, Illinois 60439*

Quantum mechanical theory disallows the model that has been used to infer the neutrino mass difference from the reported "GSI oscillations" in the rates of decay of hydrogen-like ions by electron capture. It has not been proved that the existence of mass-difference-dependent oscillations conflicts with quantum mechanics but no consistent quantum mechanical model has been shown to predict them.

PACS numbers: 23.40.Bw, 13.15.+g, 14.60.Pq, 26.65.+t

## 1. INTRODUCTION AND CONCLUSIONS

The measured electron-capture decay rates of certain hydrogen-like heavy ions have been reported to show oscillations about a decreasing exponential [1,2]. The most recently reported observed decay rates $R(t)$ were fitted [2] to

$$R(t) \propto e^{-\lambda t}\left(1 + a\cos(\omega t + \phi)\right) \tag{1}$$

where $\lambda$ is around .01 sec$^{-1}$, $\omega$ is a few times greater, and $a$ is around 0.1.

The meaning of such decay oscillations has been debated. Electron capture entails the emission of an electron neutrino, which is a coherent combination of two states in which the neutrinos have two different masses $m_1$ and $m_2$. A.N. Ivanov and P. Kienle [3] presented a model in which the oscillation frequency measures the neutrino mass-squared difference $\Delta m^2 = |m_1^2 - m_2^2|$ through

$$\Delta m^2 = 2\omega M, \tag{2}$$

where $M$ is the mass of the daughter ion. In that model, the oscillations come about through interference between different momentum components of the total wave function that depend upon the two neutrino masses. However a Comment [4] by V.V. Flambaum argued that Ref.[3] must contain some error. The total rate of decay into a channel containing a neutrino with mass $m_1$ and a channel containing a neutrino with mass $m_2$ equals the sum of the two partial rates. A fundamental principle of quantum mechanics permits no interference between the production rates of two orthogonal final states. Ivanov and Kienle, in their Reply [5] to the Comment, justified the interference by an intuitive argument in which energies and momenta are spread out in the measurement in consequence of the uncertainty principle. The interference between different values of the momentum creates the oscillations and leads to Eq.(2).

The question at issue transcends the details of the GSI experiments and it is clearly important because Eq.(2), if it is correct, enables an independent measurement of the neutrino mass difference.

It is shown below that both contentions are incorrect. The orthogonality of the two neutrino mass states does not exclude the possibility of mass-difference-dependent decay oscillations arising from interference effects but the Ivanov-Kienle model cannot describe such oscillations.

No justification in quantum mechanical theory for inferring the neutrino mass difference from an observed oscillation frequency by using Eq.(2) appears to have been put forth. It has not been proved that quantum mechanics forbids decay oscillations, but neither has it been demonstrated that quantum mechanics allows them. A solvable model could possibly resolve that question.

The time $t$ in Eq.(1) was measured by a clock for which $t = 0$ at the time of a nuclear collision in which the parent nucleus was born. A model can at most produce oscillations that depend upon the time interval between the formation and decay of the parent ion. Unless the parent ions are created either in the nuclear collision or by acquisition of a bound electron after a time shorter than the oscillation period, the oscillations will be washed out by random variations in the times of formation of the ions.

## 2. THE IVANOV-KIENLE MODEL

The Hilbert space for the decaying system consists of a part in which the parent ion is present and a part in which decay products are present. Let $\psi(t)$ be the wave function of the entire system and let $P_p$ the projection on states in which the parent nucleus is present. The decay rate $R(t)$ equals $-dS/dt$, where the survival probability $S(t)$ is given by

$$S(t) = \langle P_p \psi(t) | P_p \psi(t) \rangle \tag{3}$$
$$= \langle \psi(0) | e^{iHt} P_p e^{-iHt} | \psi(0) \rangle \tag{4}$$
$$= \int d^3\mathbf{K}\, d^3\mathbf{K}' \langle \psi(0) | \mathbf{K} \rangle \langle \mathbf{K} | e^{iHt} P_p e^{-iHt} | \mathbf{K}' \rangle \langle \mathbf{K}' | \psi(0) \rangle \tag{5}$$

$H$ is the Hamiltonian and $\mathbf{K}$ represents the total momentum. The other dynamical variables that should accompany $\mathbf{K}$ and $\mathbf{K}'$, all of them invariant under translation, have been suppressed. Those include all spins, the relative momentum of the neutrino and the daughter nucleus, and internal variables of the nucleus. $t = 0$ represents the time when the parent nucleus was created.

In the Ivanov-Kienle model, oscillations in $S(t)$ are assumed to arise only from off-diagonal parts of $\langle \mathbf{K} | e^{iHt} P_p e^{-iHt} | \mathbf{K}' \rangle$. A partly intuitive argument based on the uncertainty principle is given to show how unequal pairs $(\mathbf{K}, \mathbf{K}')$ can contribute to the decay rate and Eq.(2) is obtained. That reasoning cannot be correct. Consider first the case where there are no external fields. *H* and *P* do not commute with each other, but in the absence of external fields the total momentum operator $\hat{\mathbf{K}}$ commutes with both. Then

$$\langle \mathbf{K} | e^{iHt} P_p e^{-iHt} | \mathbf{K}' \rangle = \delta(\mathbf{K} - \mathbf{K}') \langle \mathbf{K} | e^{iHt} P_p e^{-iHt} | \mathbf{K} \rangle. \tag{6}$$

There is no interference between different values of the momentum K in Eq.(5) and no basis for Eq.(2), which relates the oscillation frequency to the neutrino mass difference, remains. This result is not surprising. For $\langle \mathbf{K} | e^{iHt} P_p e^{-iHt} | \mathbf{K}' \rangle$ not to be diagonal in **K**, the Hamiltonian must not be invariant under translation, and that implies that there is an external field.

In reality the GSI experiment was carried out in a storage ring with magnetic guide fields and additional complications introduced by electron beam cooling, and one may speculate that those will somehow induce oscillations as a result of interference between different values of the momentum. But then the oscillations will necessarily have to depend upon those external effects and their frequency may or may not depend upon the neutrino mass difference. Absent a detailed calculation, there is no basis for using Eq.(2) to determine the mass difference.

### 3. ORTHOGONALITY AND DECAY OSCILLATIONS

Flambaum [4] argued that "The final states of these reactions are different and orthogonal to each other … Therefore, the amplitudes of these reactions cannot interfere, and no oscillations are possible." It will next be proved that the mutual orthogonality of the two mass channels does not in fact imply the absence of the needed interference.

For present purposes it is not necessary to include the positron decay channel. The Hilbert space then consists of three mutually orthogonal parts, one with a parent ion and nothing else, one with a daughter nucleus and a neutrino having mass $m_1$, and one with a daughter nucleus and a neutrino having mass $m_2$. The Hamiltonian can be written as

$$H = H_0 + V \tag{7}$$

where

$$\begin{aligned} H_0 &= P_p H P_p + P_1 H P_1 + P_2 H P_2 = H_p + H_1 + H_2 \\ V &= P_p H P_1 + P_1 H P_p + P_p H P_2 + P_2 H P_p \end{aligned} \tag{8}$$

$P_p, P_1$, and $P_2$ project respectively on states containing the parent ion, a daughter ion plus a neutrino having mass $m_1$, and a daughter ion plus a neutrino having mass $m_2$. All components of *V* are proportional to the weak interaction strength *G*. The components of $H_0$ do not contain the weak interaction but they may include external fields.

The wave function $\psi(t)$ for the entire system has the form

$$\psi(t) = \varphi_p(t) + \varphi_1(t) + \varphi_2(t), \tag{9}$$

where
$$\varphi_j(t) = P_j \psi(t). \tag{10}$$

The $\varphi_j$ are functions of the dynamical variables within their respective channels. They are orthogonal to each other because they project on orthogonal channels.

From the Schroedinger equation,

$$i\dot{\varphi}_p(t) = H_p \varphi_p(t) + V_{p1}\varphi_1(t) + V_{p2}\varphi_2(t) \tag{11}$$
$$i\dot{\varphi}_1(t) = H_1 \varphi_1(t) + V_{1p}\varphi_p(t) \tag{12}$$
$$i\dot{\varphi}_2(t) = H_2 \varphi_2(t) + V_{2p}\varphi_p(t) \tag{13}$$

where
$$V_{jk} = P_j V P_k. \tag{14}$$

From Eqs.(12,13), $\varphi_1(t)$ depends upon $m_1$ through $H_1$, and similarly for $\varphi_2(t)$ and $m_2$. Then from Eq.(11), $\dot{\varphi}_p(t)$ is a sum of three terms one of which depends upon $m_1$ and one upon $m_2$. Those terms are simply added in $\dot{\varphi}_p(t)$. They are not in separate channels so they may give rise to interference effects in the survival probability $\langle \varphi_p(t) | \varphi_p(t) \rangle$, and therefore in the decay rate.

Stated otherwise, the two neutrino channels are coupled through their interaction with the parent nucleus. That coupling can cause $\varphi_2(t)$ to depend upon $m_1$ and $\varphi_1(t)$ to depend upon $m_2$. Thus $\langle \varphi_1(t) | \varphi_1(t) \rangle$ and $\langle \varphi_2(t) | \varphi_2(t) \rangle$ may each be a function of both masses and orthogonality of the two channels does not alone forbid interference effects involving the two masses. Whether such interference effects, if they do exist, can result in oscillations that depend upon the neutrino mass difference remains an open question. That possibility is not excluded by the orthogonality of the two mass channels but no acceptable theory has yet validated it.

## ACKNOWLEDGEMENTS

I thank John P. Schiffer for many critical questions and comments. This work is supported by the U.S. Department of Energy, Office of Nuclear Physics, under Contract No. DE-AC02-06CH11357.

----------------------------------